\documentclass[prd,twocolumn,amsmath,amssymb,floatfix,superscriptaddress,nofootinbib,preprintnumbers]{revtex4-1}
\pdfoutput=1

\usepackage{preamble}

\showcommentsfalse

\ifshowcomments
    {}
\else
    \renewcommand{\sout}[1]{\unskip}
\fi

\begin{document}

\title{
Mass Beyond Measure: Eccentric Searches for Black Hole Populations 
}

\author{Nicholas DePorzio}
\affiliation{Physics Department, Boston University, Boston, MA 02215, USA}
\affiliation{Department of Physics, Harvard University, 17 Oxford St, Cambridge, MA 02138, USA}
\author{Lisa Randall}
\affiliation{Department of Physics, Harvard University, 17 Oxford St, Cambridge, MA 02138, USA}
\author{Zhong-Zhi Xianyu}
\affiliation{Department of Physics, Tsinghua University, Beijing 100084, China}

\begin{abstract}

Stellar mass binary black holes of unknown formation mechanism have been observed, motivating new methods for distinguishing distinct black hole populations. This work explores how the orbital eccentricity of stellar mass binary black holes is a viable conduit for making such distinctions. Four different production mechanisms, and their corresponding eccentricity distributions, are studied in the context of an experimental landscape composed of mHz (LISA), dHz (DECIGO), and Hz (LIGO) range gravitational wave detectors. We expand on prior work considering these effects at fixed population eccentricity. We show that a strong signal corresponding to subsets of eccentric populations is effectively hidden from the mHz and dHz range gravitational wave detectors without the incorporation of high eccentricity waveform templates. Even with sufficiently large eccentricity templates, we find dHz range experiments with a LISA-like level of sensitivity are unlikely to aid in distinguishing different populations. We consider the degree to which a mHz range detector like LISA can differentiate among black hole populations independently and in concert with follow-up merger detection for binaries coalescing within a 10 year period. We find that mHz range detectors, with only $e < 0.01$ (nearly circular) sensitivity, can successfully discern eccentric sub-populations except when attempting to distinguish very low eccentricity distributions. In these cases where $e < 0.01$ sensitivity is insufficient, we find that the increase in event counts resulting from $e < 0.1$ sensitivity provides a statistically significant signal for discerning even these low eccentricity sub-populations. While improvements offered by $e<0.1$ sensitivity can generally be increased by $\mathcal{O}(1)$ factors with $e<0.4$ sensitivity, going beyond this in eccentricity sensitivity provides negligible enhancement.

\end{abstract}

\maketitle
\onecolumngrid
\hrule
\tableofcontents
\vspace{20pt}
\hrule
\vspace{20pt}
\twocolumngrid

\section{Introduction}

The initial detection of stellar mass binary black hole pair (BBH) mergers by LIGO/VIRGO ~\cite{LIGOScientific:2018mvr}, since supplemented by numerous similar observations ~\cite{KAGRA:2021duu, LIGOScientific:2021djp}, has ushered in an era of gravitational wave astronomy and suggests the existence of a sizeable BBH abundance. Our understanding of early Universe physics, the nature of dark matter, and especially astrophysical black hole production all stand to benefit from this data and must respond to the challenges presented by observed black holes of anomalous mass ~\cite{LIGOScientific:2020iuh}. For these reasons, it is both timely and compelling to consider means by which we can infer the formation channels of the black holes we observe with gravitational wave detectors. 

Various properties have been proposed as useful probes of the formation mechanisms of black holes, including black hole spin and mass ~\cite{Nishizawa:2016jji, Franciolini:2021xbq,McIsaac:2023ijd}. This work investigates how the orbital eccentricity of BBH pairs can offer insight into black hole formation channels ~\cite{Breivik:2016ddj, Armitage:2005xq, Amaro-Seoane:2009iyc, Amaro-Seoane:2009xmw, Nishizawa:2016eza,Antonini:2012ad, Antonini:2015zsa, Baruteau:2010sm}. Prior work has demonstrated that various static and dynamic processes can lead to populations of black holes with different distributions in their orbital eccentricity  ~\cite{Nishizawa:2016eza, Rodriguez:2018pss, Randall:2018qna, Fang:2019dnh}. As a typical BBH approaches merger, orbital eccentricity is quickly attenuated through the radiation of gravitational waves, with little residual eccentricity at merger ($\gtrsim 10^2$ Hz) frequencies for stellar mass black holes. This process does not exclude the possibility of a BBH having considerable eccentricity at lower frequencies, including the mHz range probed by detectors like LISA ~\cite{LISA:2022yao}. This eccentricity attenuation is also affected by the BBH environment, where dense environments allow the possibility of BBH pairs maintaining their eccentricity through Kozai-Lidov oscillations ~\cite{Randall:2019sab, Naoz_2016, Chandramouli:2021kts, Hoang:2019kye}. These dynamics introduce the interesting possibility of exploring BBH formation channels through a multi-frequency observation campaign composed of detectors like LIGO  and VIRGO exploring the near-merger regime, and lower frequency detectors like LISA exploring the regime far from merger.

Prior work has considered the observable consequences of eccentricity in mHz frequency detectors~\cite{Randall:2021xjy,Randall:2019znp}. Amongst such effects are an increase in the perceived number of events in a given frequency window, and a suppression to the detector signal-to-noise ratio (SNR). While these works have demonstrated the detection response for any single eccentricity of BBH, the effects of a BBH population possessing a distribution in eccentricity have not yet been studied. In this work, we consider the observational consequences of introducing a distribution in eccentricity to the BBH population. We consider four different single-peaked distributions and mixtures of those distributions in varying ratios.  

In Section \ref{section2}, we introduce the dynamics of an eccentric BBH pair. In Section \ref{section3} we review the effects of introducing eccentricity on the detection signal and sensitivity. We also present the expected observation in mHz and dHz range detectors for eccentric BBH populations. In Section \ref{section4} we explore how experimental sensitivity to different eccentricities impacts observational degeneracies between different BBH distributions. In Section \ref{section5}, we consider whether a mHz range detector such as LISA can significantly distinguish populations of BBH pairs. We conclude in Section \ref{section6}.

\section{Dynamics of Binary Black Hole Inspiral} 
\label{section2}

Only for sufficiently large metric perturbations is a higher-order general relativistic description necessary to model the gravitational signature of a BBH. For mHz range gravitational wave signals, typical binary systems are far from merger and such considerations are not necessary. To this effect, we shall only consider the quadrupole gravitational wave emission of the signal, modeled to the leading post-Newtonian order in General Relativity. We expect this approximation to break down in the near-merger regime. The extent that we consider the observation of merger events in this work is limited only to the calculation of the merger time. We assume the merger time approximation we utilize is sufficiently accurate for the purposes of this study, though acknowledge that corrections would be introduced by a more careful consideration.

We describe a BBH, composed of masses $m_1$ and $m_2$, in terms of its total mass

\seteq{mtotal}{m \equiv m_1+m_2{\rm ,}}
its reduced mass 

\seteq{mreduced}{\mu \equiv \frac{m_1 m_2}{m}{\rm ,}}
its chirp mass

\seteq{mchirp}{m_c \equiv \frac{\mu^{3/5}}{m^{2/5}}{\rm ,}}
the semi-major axis of its orbit, $a$, and the orbital eccentricity, $e$. A non-circular binary with $e>0$ will tend to circularize through emission of higher harmonic quadrupole radiation. To the post-Newtonian level, these dynamics are described by the Peters' equations ~\cite{Peters:1964zz}

\seteq{peters1}{\frac{da}{dt} = - \frac{64}{5}\frac{G^3 \mu m^2}{c^5 a^3} \frac{1+\frac{73}{24}e^2 + \frac{37}{96}e^4}{(1-e^2)^{7/2}}{\rm ,}}

\seteq{peters2}{\frac{de}{dt} = - \frac{304}{15}\frac{G^3 \mu m^2}{c^5 a^4} \frac{e(1+\frac{121}{304}e^2)}{(1-e^2)^{5/2}}\eqstop}
This set of equations fixes the relationship $a(e)$ in the evolution of a binary system. Further, we can describe the peak frequency of quadrupole emission by ~\cite{Randall:2017jop,Randall:2021xjy}
\seteq{fp}{f_p \approx \frac{\sqrt{Gm}(1+e)^\gamma}{\pi (a(1-e^2))^{3/2}]}{\rm ,}}
with $\gamma = 1.1954$, which departs from the $e=0$ relationship $f_p = 2 f_{\rm orb}$ as eccentricity increases. While emission also occurs at higher harmonics, in this work we  consider only the signal of a single BBH to be comprised of the emission at $f_{p}$ - a more detailed analysis might consider the expected enhancement due to simultaneous observation of the emission at multiple frequencies.

Ignoring corrections which occur near the merger, ~\refeq{peters1} and \refeq{peters2} can be used to estimate the time to merger by evolving until $a$ is of order the black hole radius ~\cite{Randall:2019sab}
\begin{equation}
\label{eq:tmerge}
    \begin{aligned}
        \begin{array}{l}
            t_{\rm merge} \approx  \frac{5}{256}\frac{c^5}{\mu m^2 G^3} \\
            \phantom{xxxxxxxx}\times  \left(\frac{(1+e)^{2\gamma/3} G^{1/3}m^{1/3}}{(1-e^2) (f_p \pi)^{2/3}}\right)^4 (1-e^2)^{7/2}\eqstop{} 
        \end{array}
    \end{aligned} 
\end{equation}
At fixed peak emission frequency, $f_p$, the lifetime given by \refeq{tmerge} grows monotonically with $e$. This implies that at a given frequency, a more eccentric binary will take longer to merge than a less eccentric binary radiating at the same frequency.

\begin{figure}[t]
    \centering
    \includegraphics[width = \linewidth]{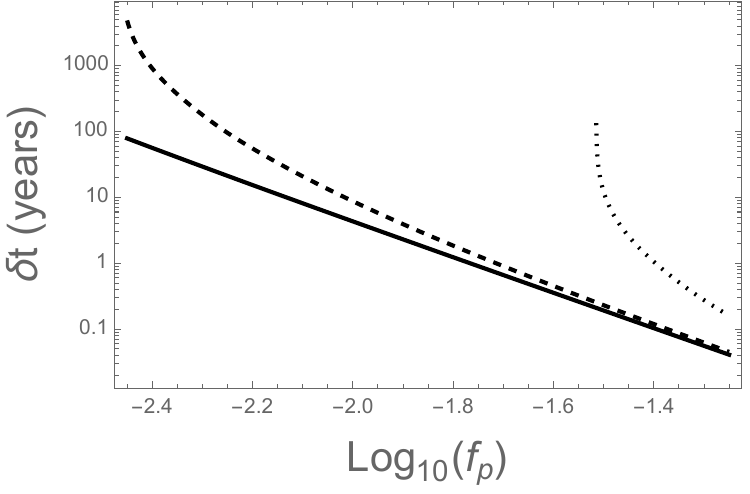}
    \caption{The coalescence time for a merging BBH pair for three choices of $e_*$, the eccentricity at reference frequency $f_* \equiv 10$ Hz. The solid line corresponds to $e_*=10^{-5}$, the dashed line to $e_* = 10^{-4}$ and the dotted line to $e_* = 10^{-3}$. More eccentric binaries spend more time at a particular frequency in the course of their evolution.}
    \label{fig:deltat_of_fp}
\end{figure}

Combining \refeq{peters1} and \refeq{peters2} with \refeq{fp}, we see that specifying the mass, peak frequency, and eccentricity of a binary at a fixed point in time determines the evolution of its physical properties at all times. Alternatively, we can use these same relations to describe the frequency evolution (i.e. chirping) of the BBH with time 

\seteq{chirping}{\frac{dt}{df_p} = \frac{5c^5}{96\pi^{8/3}} (G m_c)^{-5/3} f_p^{-11/3} \mathcal{F}(e){\rm ,}} 
\begin{equation}
\label{eq:scriptF1}
    \begin{aligned}
        \begin{array}{cc}
            \mathcal{F}(e) \equiv &  \frac{(1+e)^{8\gamma/3 - 1/2}}{(1-e)^{3/2}} \big( (1+e)(1+\frac{7}{8}e^2) \\
             & -\frac{\gamma}{288}e(304 + 121e^2) \big)^{-1}\eqstop{}
        \end{array}
    \end{aligned} 
\end{equation}

\begin{figure}[t]
    \centering
    \includegraphics[width = 1.0\linewidth]{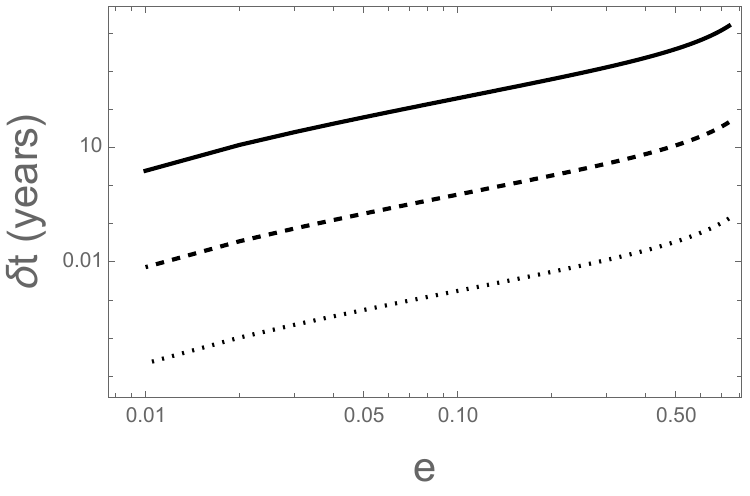}
    \caption{Same as ~\reffig{deltat_of_fp}, but in terms of eccentricity. More eccentric binaries spend less time at a particular eccentricity in the course of their evolution.}
    \label{fig:deltat_of_e}
\end{figure}

Noting that $\mathcal{F}(e)\rightarrow 1$ as $e\rightarrow 0$, we see that the $\mathcal{F}(e)$ acts as a suppression to the chirping ($df_p/dt$) of a circular BBH pair.  The amount of time a BBH spends at a given frequency, $\delta t$, as a function of peak emission frequency is shown in \reffig{deltat_of_fp}, and similarly in terms of eccentricity in \reffig{deltat_of_e} for three different choices of $e_* \equiv e(f_p=10~{\rm Hz})$, the BBH eccentricity at a reference frequency of 10 Hz.

\begin{figure*}
  \centering
  \includegraphics[width=\textwidth]{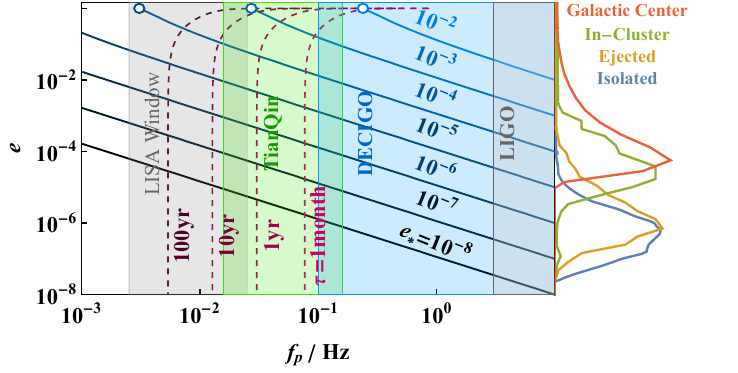}
  \caption{Evolution of black hole binary eccentricity as a function of peak quadrupole emission frequency. The evolution for several choices of $e_* \equiv e(f_p=10 {\rm ~Hz})$ is shown - for each, the lower limit of $f_p$ is indicated by a circle. Points to the right of the shown magenta lines will merge within the indicated time period. Along the right edge, drawn at $f_p = 10~{\rm Hz}$, are the $e_*$ distributions corresponding to four different formation channels. }
  \label{fig:edist_evolution}
\end{figure*}

In \reffig{edist_evolution} we demonstrate the evolution of $e(f_p)$ for various choices of $e_*$. Note that for any choice of $e_*$ there will exist a lower bound on $f_p$ corresponding to $e=1$. As an example, it is worth noting that sufficiently eccentric BBHs with $e_* \gtrsim 10^{-3}$ will never radiate in the LISA window but will still merge in the LIGO window; in such a situation, the difference on BBH counts in LISA versus LIGO might signal the presence of a population of highly eccentric binaries. Likewise, while all BBHs with $e_* \lesssim 10^{-3}$ will produce a signal in LISA and eventually enter the LIGO band, only those which initially appear in the LISA band with sufficiently high frequency will make this transition in a period of time reasonable for a follow up observation in LIGO. The magenta bands in  \reffig{edist_evolution} indicate this lower bound on $f_p$ for various choices of follow-up time. 

In this work, we are concerned with the evolution of BBHs for a population distributed in  $e_*$. In \reffig{edist_evolution}, the upper limit of $f_p$ is drawn at 10 Hz, the same frequency at which $e_*$ is defined. By drawing the $e_*$ distribution at 10 Hz and tracing lines of constant $e_*$ to the left of the distribution, we gain a qualitative understanding of how various $e_*$ populations evolve through this detector landscape. We show predictions for the $e_*$ distributions generated via a sampling of astrophysical mechanisms (Isolated, Ejected, In-Cluster, Galactic Center) ~\cite{Nishizawa:2016eza, Rodriguez:2018pss, Randall:2018qna}. 

As an example, consider the evolution of an Isolated distribution, such as might describe a black hole population existing in the field of a galaxy. In this case, we see that the most likely eccentricity of the population is $e_* \approx 10^{-6}$; Such binaries will radiate in both the LISA and LIGO bands; and will merge within 10 years for binaries appearing in LISA with $f_p \gtrsim 0.02$ Hz.     

\section{Observation of Eccentric Binary Populations}
\label{section3}

We wish to consider the role that eccentricity plays in modifying a  stellar mass BBH signal entering a gravitational wave detector, as well as in the noise response of the detector to such a signal. We begin by noting there are a number of population characteristics that we assume are independent of the $e_*$ distribution. First, we assume the case of a static Universe to avoid redshifting of gravitational wave frequencies and dependence of the binary chirp mass on the expansion of a cosmological background. Beyond what we describe above, we do not incorporate higher order relativistic corrections (e.g. doppler shifting) which are expected to be more important for very eccentric events and different mass and frequency regimes than we focus on. While degeneracies between eccentricity and precession are relevant for Hz range detectors, mHz range detectors can observe many orbital periods and are less susceptible to this effect. We  also assume that the distributions in spatial position and mass do not correlate with the eccentricity distribution of the population. We note, however, that the same mechanisms that generate the eccentricity distributions of interest may also modify these distributions. For example, we do not expect populations produced by astrophysical processes near the galactic center to populate many BBH pairs at nearly extragalactic distances. Such considerations would certainly be necessary to draw definitive conclusions about the underlying formation channels, but as a step towards this goal we describe the uniform spatial distribution of BBHs by 
\seteq{rdist}{p(r) \propto 4\pi r^2 {\rm , }}
and we utilize the mass distribution inferred by LIGO to describe a stellar mass BBH population independent of eccentricity ~\cite{LIGOScientific:2018jsj} 
\setfig{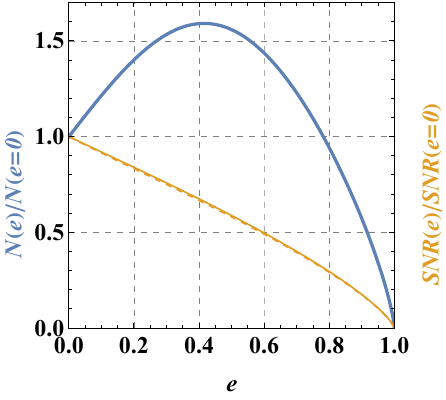}{The effect on number density and signal-to-noise ratio as a function of binary eccentricity at fixed frequency. Shown in blue is the enhancement to the observed number density of binaries in LISA relative to circular binaries. Shown in yellow (dashed) is the suppression to the LISA (DECIGO) SNR relative to circular binaries. The nearly identical effect on SNR in LISA/DECIGO indicates that suppression is driven by the change in signal rather than difference in the detector noise strains.}{fig2} 
\seteq{mdist}{p(m_1) \propto m_1^{-2.3} \eqstop}

The observed merger rate and experimental sensitivity is used by LIGO to predict the total local BBH merger rate,  
\seteq{R}{\mathcal{R} = \frac{dn}{dt} = 53.2 {\rm ~Gpc}^{-3}{\rm ~yr}^{-1} \eqstop}
which is assumed to be roughly constant for redshifts $z \lesssim 1$ ~\cite{LIGOScientific:2018jsj}. 
This rate sets the overall normalization of our distributions - given probability distributions $\mathcal{P}(e_*), \mathcal{P}(m_c), \mathcal{P}(r), \mathcal{P}(f_p)$, we can calculate the number of binaries from the distribution which will merge in a given time period.

The first place that eccentricity enters the signal is in the $f_p$ distribution of binaries. As we are always concerned with the number of binaries observed over fixed time intervals, the likelihood of a BBH signal having a particular $f_p$ upon entering the detector is given by \refeq{chirping} 
\seteq{pfp1}{p(f_p) \propto \frac{dt}{df_p} \eqstop}
The only remaining population parameter is $e_*$, whose relic distribution $p(e_*)$ will be provided by one of the underlying formation channels shown in \reffig{edist_evolution}. 

Combining these distributions provides a measure of the expected number of events in a given time interval
\seteq{Ncounts}{N_{\rm events} = \int \mathcal{R}p(m_c)p(r)p(f_p, e(f_p))p(e_*)dm_c dr df_p de_*}
It is apparent that eccentricity affects our signal, the frequency binned event count, in three ways: (1) $dt/df_p$ experiences the competing effects of an enhancement through \refeq{scriptF1} and a change in the $\partial f_p / \partial e$ evolution, (2) $f_p(t)$ is modified to satisfy $e(f_p=10 \rm{~Hz}) = e_*$, and (3) the likelihood of a particular $e_*$ is weighted by $p(e_*)$. 

We adopt a simplified model for the SNR in the presence of chirping binaries 

\seteq{rho1}{\varrho(f_p, e)^2 = 4 \int dt \frac{h_c^2(f_p(t), e=0)}{S_N(f_p(t))} (1-e(t))^{3/2}} 

shown to be a good approximation ~\cite{Randall:2019znp} to the true SNR 

\seteq{rho2}{\varrho^2 = 4 \sum_n \int dt \frac{h_n^2(f_n(t))}{S_N(f_n(t))} }
which involves a summation over the $n$ harmonic components of the gravitational wave emission. Here, $S_N(f)$ is the noise strain and $h_c^2 = \Sigma h_n^2$ is the signal strain averaged over an orbital period. Noting that $\varrho \propto (1-e)^{3/4}$ also scales inversely with $r$, we see that there should be a suppression to the sensitive volume and expected number of counts with respect to the circular case given by $(1-e)^{9/4}$. The product of this suppression term and the enhancement due to $\mathcal{F}(e)$ yields an overall effect to the event count with respect to circular BBHs that results from introducing eccentricity to  BBHs at fixed $f_p$ (non-chirping) shown in \reffig{fig2}. We also show the suppression to the SNR for  both LISA and DECIGO to demonstrate that this effect is driven by the $f_p(e)$ dynamics rather than the specific shape of the detector noise curve. 

\begin{figure*}[ht]
  \centering
  \includegraphics[width=\textwidth]{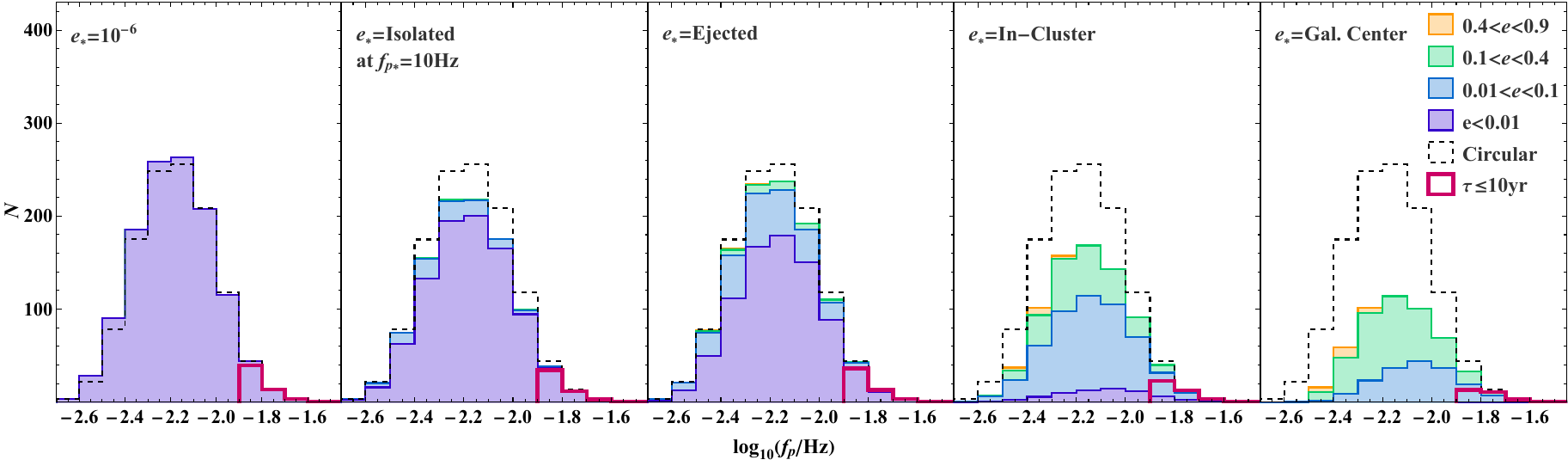}
  \caption{Number of observed counts in LISA for fixed, Isolated, Ejected, In-Cluster, and Galactic Center $e_*$ distributions. Dashed line indicates expected counts for a perfectly circular distribution. While a fixed $e_*$ distribution would exhibit a lower bound in $f_p$, distributions in $e_*$ distribute counts across frequency bins for all choices of $e_{\rm cut}$. Magenta lines indicate events which will merge within 10 years and have $e_{\rm cut}=0.9$, suggesting events which can have an observable Hz range terrestrial observation follow-up.}
  \label{fig:edist_counts}
\end{figure*}
\begin{figure*}[htb]
  \centering
  \includegraphics[width=\textwidth]{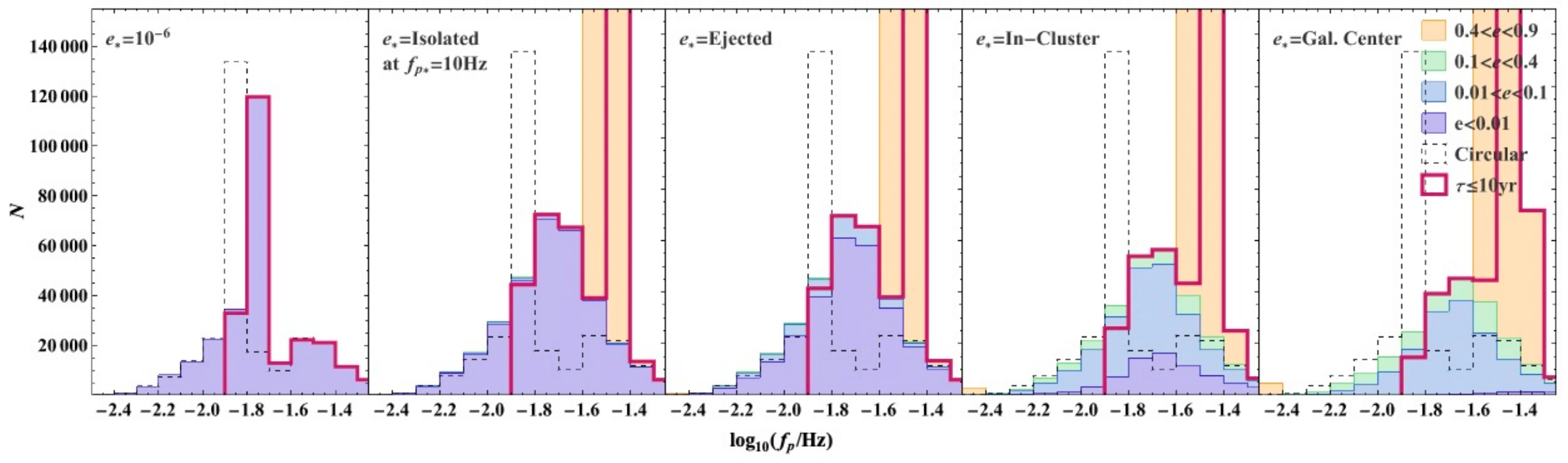}
  \caption{Same as \reffig{edist_counts} except using DECIGO noise curve. Note that no redshift effects have been included.}
  \label{fig:edist_counts_decigo}
\end{figure*}

Our detection ability will also be influenced by which eccentricities we assume our detectors are able to observe - this is determined by the degree of eccentricity gravitational waveform templates have been developed for. This will introduce a cut into our event count over the parameter space wherever the eccentricity is larger than the template upper bound on eccentricity

\seteq{thetaecut}{\Theta_{e_{\rm cut}}(e) = \Theta(e_{\rm cut}-e)\eqstop{}}
Likewise, the calculated SNR will determine whether a signal is observable. We impose SNR $> 8$ over 10 years of LISA observation as the criteria by which to consider events observable.
\seteq{thetasnr}{\Theta_{\rm SNR}(r, f_p, m_c, e_*) = \Theta(\varrho(r, f_p, m_c, e_*) - 8)\eqstop{}}
Combining these terms, the observable number count is given by 
\begin{equation}
\label{eq:scriptF2}
    \begin{aligned}
        \begin{array}{cc}
             N_{\rm events} =  &\int \mathcal{R}p(m_c)p(r)p(f_p, e(f_p))p(e_*) \\
             \phantom{xxx}&\times \Theta_{\rm SNR}\Theta_{e_{\rm cut}}dm_c dr df_p de_*\eqstop{}
        \end{array}
    \end{aligned} 
\end{equation}

In the case of a fixed eccentricity population, there is a mininum $f_p$ below which no gravitational wave emission exists. Likewise, there will be a lower bound on frequency imposed by the maximum $e$ with available templates. Transitioning from fixed $e_*$ to a distribution, there is no longer a single lower bound on $f_p$ as each choice of $e_*$ will produce a different bound. Sensitivity to the shape of the $e_*$ distribution is conveyed by how the counts per frequency bin shift as we move $e_{\rm cut}$. This can be seen, for example, in the shifting of the Galactic Center distribution counts in \reffig{edist_counts}.

In \reffig{edist_counts} we show the expected number of observable events in LISA over a 10 year observation period for five choices of the BBH eccentricity distribution. In each case, we also highlight the subset of events which will merge within 10 years as an example of candidates that could potentially be followed-up with a LIGO-like detection. For reference, we compare with what the expected number of counts would be assuming a BBH population of only circular binaries, denoted by the black dashed line. In the leftmost panel, we show the result for a fixed - and relatively small - eccentricity to indicate the close matching with the circular case. Progressing to the right, we consider the Isolated distribution, which is peaked at $e_* \approx 10^{-6}$ but extends to both higher and lower eccentricities. As indicated by the four different colored histograms, we begin to lose sensitivity to events as we impose progressively lower bounds on $e_{\rm cut}$. In one case, no events would be observed for the Galactic Center distribution without eccentricity templates of $e_{\rm cut} \geq 0.01$. In this situation, as demonstrated in ~\cite{Randall:2019znp}, a mismatch between the LIGO merger event rate and LISA merger event rate would suggest the existence of an eccentric sub-population, despite having no direct observations in the LISA detector. This lack of observation in the LISA detector can therefore itself be considered a signal of eccentric BBH populations.

Because the low frequency signal is suppressed, it is of interest to see what sensitivity an intermediate range (dHz) detector might contribute. As an example we consider the DECIGO noise curve ~\cite{Yagi:2011wg}. In ~\reffig{edist_counts_decigo}, we show the expected number of event counts in DECIGO over a 10 year observation period. Considering we explore the same mass distribution, the overall increase in events with respect to the LISA case is driven by the very large enhancement in detector sensitivity, mitigated by a suppression in $dt/df_p$ due to the higher frequencies being considered. Below, we consider how the observational power of DECIGO to these events is modified by weakening the DECIGO noise curve. 

In ~\reffig{edist_counts_decigo} we note the emergence of a second cluster of high frequency events at very high eccentricities ($0.4 < e < 0.9$). To understand the origin of this feature, we can refer back to ~\reffig{deltat_of_fp}; highly eccentric BBH pairs in any distribution will radiate only above their cutoff frequency, and when radiating near the cutoff frequency they will be at their highest eccentricities as well as spend much longer time at those frequencies with respect to less eccentric counterparts. Because the detector SNR is a response to integrating a BBH signal over time, it will respond favorably to signals which remain near the peak of their sensitivity curve for longer durations. This offers an explanation for the general shape of these distributions - at low frequencies, the detector sensitivity is too weak, whereas at higher frequencies the integration time will be too limited \textit{unless} the binary pair is sufficiently eccentric. This interplay produces a hidden cluster of event counts for any distribution which contains extremely high eccentricity BBH pairs in its distribution that will not be observed unless very high eccentricity waveform templates are available to the detector. We present this result with the caveat that the effects of redshift on the SNR for dHz detectors are important, as is the redshift dependence of the BBH distributions we consider, but we do not include them in this simplified analysis. For this reason, ~\reffig{edist_counts_decigo} should only serve as a conceptual illustration.

Note that the DECIGO noise curve presents a $\mathcal{O}(10^4)$ improvement in the root power spectral density sensitivity over the LISA detector at peak. We consider what changes the more pessimistic case of a DECIGO-like detector, which possesses the same shape for the noise curve of DECIGO but is manually suppressed by an overall factor which aligns the minimum of its noise curve to be at the same sensitivity level as the minimum of the LISA noise curve. Such a DECIGO-like detector allows us to consider a case where a detector like LISA is constructed at the intermediate frequency (dHz) range of the DECIGO detector. We find this DECIGO-like detector observes zero BBH pairs, in all frequency bins, and for all cuts in the eccentricity template. In this case it appears the weakened sensitivity no longer sufficient to overcome the small time-integrated SNR at these frequencies. We conclude that intermediate frequency (dHz) detectors are of limited use in searching for stellar mass binary black hole populations, unless they posses several orders of magnitude sensitivity improvements over modern designs for mHz range detectors, such as LISA. For this reason, we will not incorporate results for dHz range detectors in the following sections. We emphasize that dHz range detectors may remain useful for other searches, such as those relating to intermediate mass black holes ~\cite{Sedda:2021yhn, LIGOScientific:2017zid}. 

\section{Using Eccentricity to Break Observational Degeneracies} 
\label{section4}

Considering the Isolated and Ejected panels in \reffig{edist_counts}, notice that there is a similar number of events at each frequency bin, as expected due to the similarity of their distributions in \reffig{edist_evolution} - that is, there is a significant amount of degeneracy between these two models in the dataset. It is useful to ask how such degeneracies in the LISA dataset can be broken through the use of choices in $e_{\rm cut}$. To answer this question, we consider a model for the distribution of $e_*$ given by 

\begin{equation}
\label{eq:festar1}
    \begin{aligned}
        \begin{array}{cl}
            f(e_*) = & \phantom{+} A_{\rm Isolated} f_{\rm Isolated}(e_*) \\
            & + A_{\rm Ejected} f_{\rm Ejected}(e_*)  \\
            & + A_{\rm In-Cluster} f_{\rm In-Cluster}(e_*) \\
            & + A_{\rm Gal. Center} f_{\rm Gal. Center}(e_*) {, }
        \end{array}
    \end{aligned} 
\end{equation}

such that 
\seteq{distsum1}{\sum A_i =1 \eqstop}

\begin{figure*}[hp!]
  \centering
  \includegraphics[width=\textwidth]{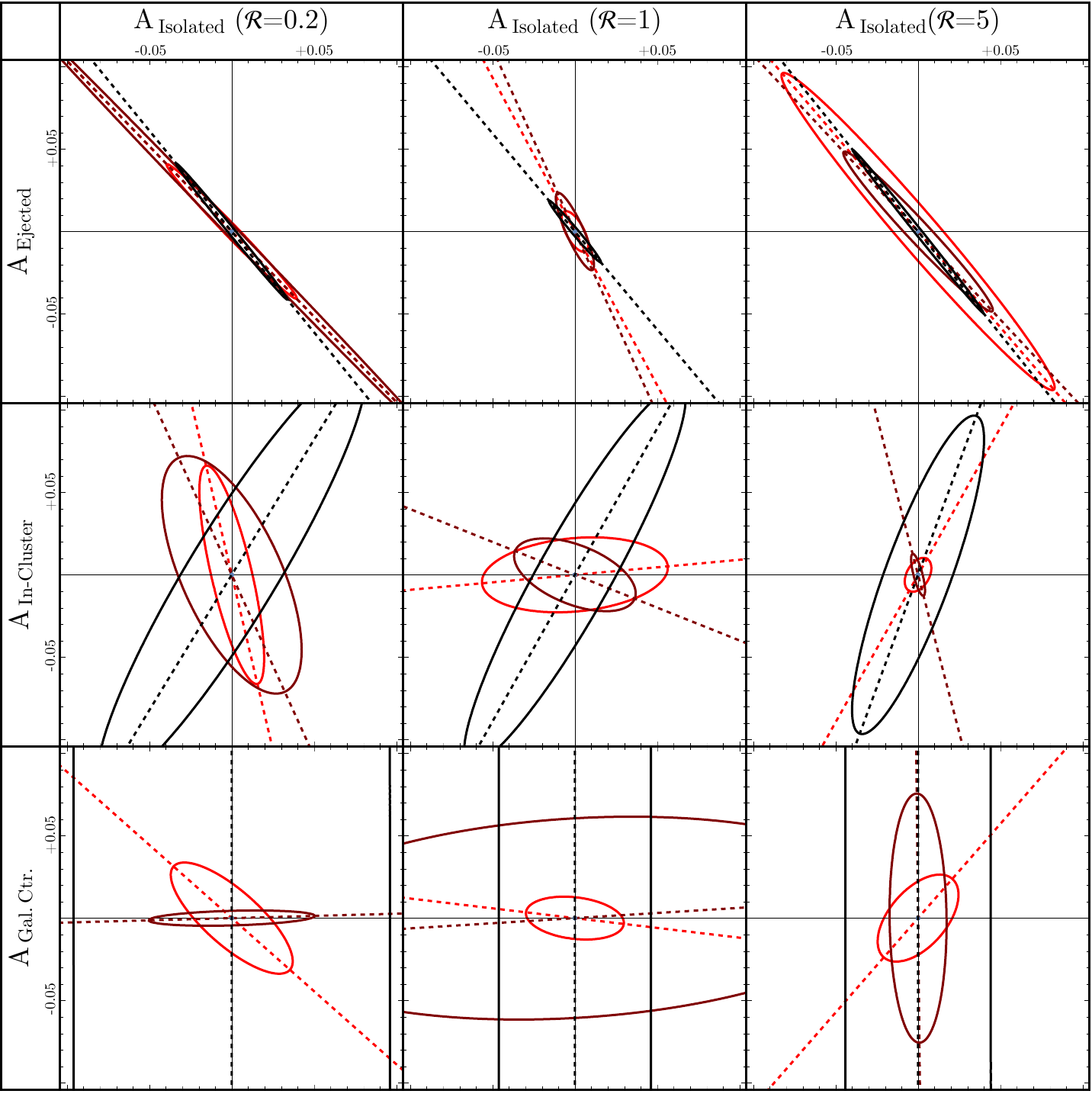}
  \caption{
    Forecasted 3$\sigma$ constraint contours on coefficients of binary $e_*$ distributions for a LISA-like experiment. In all cases, $\mathcal{R} = A_{\rm isolated}/A_j$ where $A_j$ represents the abundance of the corresponding $e_*$ distribution indicated at the left of the figure. In each panel, only two $e_*$ distributions are considered at a time with $A_{\rm isolated} + A_j = 1$. Three different abundance ratios $\mathcal{R} \equiv A_{\rm isolated}/A_j$ are considered, as well as three choices of maximum observable eccentricity $e_{\rm cut}= 0.1$ (black), $0.4$ (dark red), and $1.0$ (light red). All panels span $\pm 0.1$ from the center point. Generally, ellipse size indicates constraining power, while ellipse angle indicates parameter degeneracy (e.g. the Isolated and Ejected distributions are highly degenerate). 
  }
  \label{fig:contours}
\end{figure*}

To quantify the observational degeneracy between models in our dataset, we  construct a Fisher information matrix for the two parameter model $A_{\rm Isolated}$, $A_j$ where $j$ corresponds to one of the Ejected, In-Cluster, or Galactic Center distributions ~\cite{Gair:2022fsj}. The four elements of the corresponding Fisher matrix are defined by
\seteq{fisher}{F_{ij} \equiv \sum_{f_{p,k}}\sqrt{N_k} \frac{\partial N_k}{\partial A_i}\frac{\partial N_k}{\partial A_j} {, }}
where the index $k$ runs across each of the $f_p$ bins. We derive three separate $F_{ij}$, one for each of $e_{\rm cut} = 1.0, 0.4, 0.1$. For each of these, we consider three different fiducial choices of the ratio
\seteq{Rscript1}{\mathcal{R} \equiv \frac{A_{\rm Isolated}}{A_j} {, }}
which indicates the relative abundance of the two populations being considered. 
In total, this procedure yields nine versions of each $F_{ij}$ for each choice of two distributions. Using the Fisher information, we will consider how the degeneracy of the Isolated distribution with the other more eccentric distributions is modified as we explore the dataset at different eccentricity cutoffs. We show the resulting covariance contours between $A_{\rm Isolated}$ and $A_j$ in \reffig{contours}. 

\begin{figure*}[ht]
  \centering
  \includegraphics[width=\textwidth]{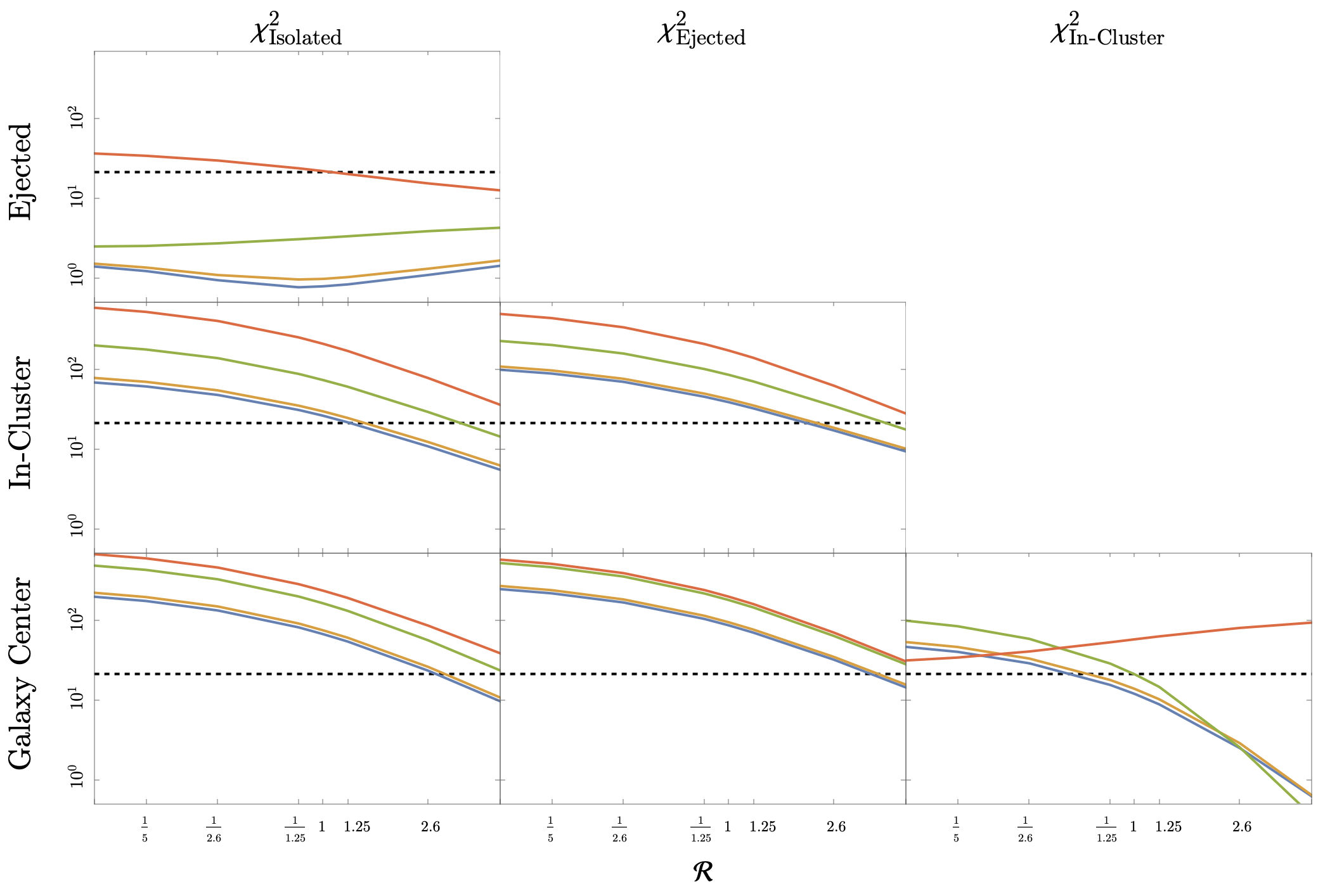}
  \caption{
  The $\chi^2$ values corresponding to LISA measurements assuming a model where the distribution indicated by the top axis comprises 100\% of the BBH distribution (the null hypothesis, $H_0$) compared against an observed dataset comprised of a test BBH distribution, indicated by the left axis, combined with the model distribution in a ratio given by $\mathcal{R}$. The value of $\mathcal{R}$ corresponds to the ratio of model distribution to test distribution (i.e. large $\mathcal{R}$ is mostly comprised of the model distribution).  Plotted for $\mathcal{R} \in [0.125, 5.0]$. Here, the dataset is the absolute counts per frequency bin in LISA for a choice of distribution and $e_{\rm cut}$. Shown are four choices of $e_{\rm cut} = 0.9$ (blue), $0.4$ (orange), $0.1$ (green), $0.01$ (red). The black dashed line indicates the $\chi^2$ value corresponding to a $p$-value of 0.05.
  }
  \label{fig:chi2}
\end{figure*}

We might expect that $e_{\rm cut}=1$, containing the most counts, would always offer the greatest constraining power. One example of an exception to this expectation is for the In-Cluster, $\mathcal{R}=1$ case. Note that while in general adding more event counts will help to minimize error, there are cases where  having sensitivity to all eccentricities - without the ability to differentiate the eccentricity of individual events - results in a total event counting between two distributions that is more similar, and less discerning, than what would have been produced by imposing an eccentricity cutoff. That is, while increasing $e_{\rm cut}$ results in more events being observed, which reduces error, it may also generate a signal that is more similar to the other distributions, reducing their distinguishability. This example illustrates that there is a distinction between the ability to observe up to a given eccentricity (giving the total events in a frequency bin with $e < e_{\rm cut}$), and the ability to actually measure the eccentricity of an observed event (i.e. tagging the value of $e_*$ for each event in a frequency bin with $e<e_{\rm cut}$); the advantage of the latter case is that it enables an experiment to bin counts by $e_*$ to provide better constraining power. An approximation to such a procedure would resemble combining the contours of any single panel in \reffig{contours} for a more precise overall result.\footnote{Note that this is only approximately true as binning by $e_*$ would change the total events contributing to a single contour.}
  
We note that for all choices of $\mathcal{R}$, \reffig{contours} indicates that the angle between contours when we compare against the more eccentric distributions (In-Cluster, Galaxy Center) become much more pronounced. This suggests that the ability to measure the value of $e_*$ in BBH events with LISA will break observational degeneracies when both low and high eccentricity populations are present.

\section{Distinguishing Black Hole Production Channels with Gravitational Wave Detectors} 
\label{section5}

\begin{figure*}[ht]
  \centering
  \includegraphics[width=\textwidth]{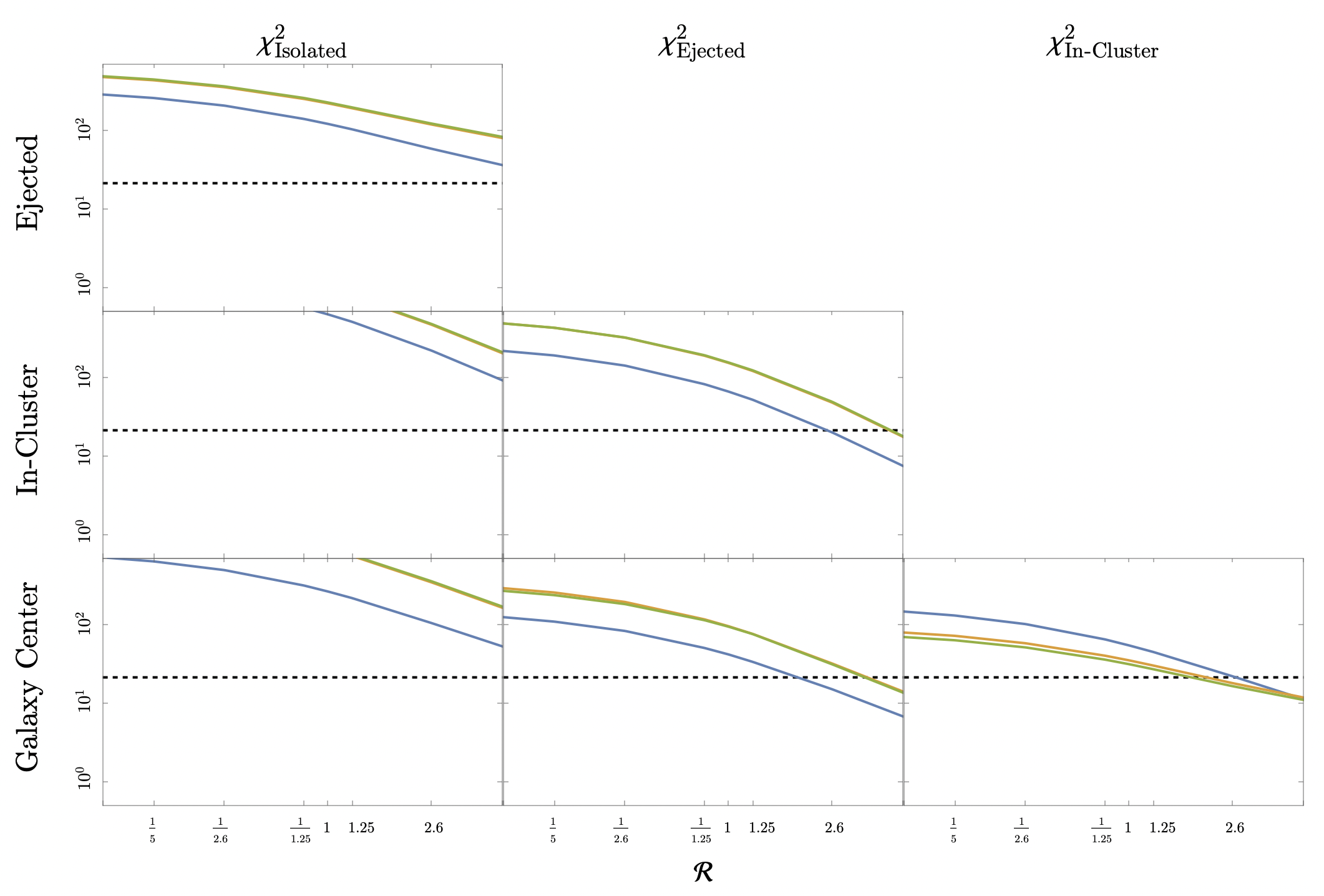}
  \caption{
  Same as \reffig{chi2} execept that here, the dataset is the difference in counts per frequency bin in LISA  between an $e_{\rm cut}=0.01$ observation and either an $e_{\rm cut}=0.1$ (blue), $e_{\rm cut}=0.4$ (orange), or $e_{\rm cut}=0.9$ (green) observation. 
  }
  \label{fig:chi2_second}
\end{figure*}

We now wish to consider how the dataset sensitivity to the population coefficients explored in Sec. \ref{section4} translates to the significance at which an experiment like LISA might be able to distinguish the BBH population deriving from different eccentricity distributions. We pose the question - assuming one $e_*$ distribution, how much of the BBH population needs to be derived from a different distribution before LISA can rule out the presence of the assumed distribution with statistical significance. We emphasize that our intention in this work is  to consider a highly simplified statistical analysis of the ability of mHz range detectors to distinguish BBH populations. We will not consider effects such as covariance between data bins, the statistics of combining datasets, experimental uncertainty on measured eccentricity (i.e. resolving an event to the wrong eccentricity bin), or a more detailed accounting of detector error as might be better explored through a more comprehensive bayesian analysis ~\cite{Digman:2022igm}. We pose as the null hypothesis, $H_{0}$, that the observed dataset is generated by an underlying distribution of BBH pairs, where we sequentially consider each of the four distributions for the null hypothesis. In all cases, the alternative hypothesis, $H_{1}$, is that the observed dataset is not drawn from the $H_0$ BBH distribution. We then assume the dataset generated by a linear combination of the $H_0$ distribution and, sequentially, each of the other distributions as the observation and ask if the observed dataset is sufficient to reject $H_0$ at 95\% significance. Here, we will again quantify the portion of the population derived from the $H_0$ distribution by the parameter $\mathcal{R}$ - large values of $\mathcal{R}$ corresponding to a greater portion of the population being drawn from the $H_0$ distribution. We then perform the \textit{Pearson chi-square goodness-of-fit test} and construct the corresponding statistic for the dataset: 

\seteq{chisquare1}{\chi^2 = \sum_{f_{p,i}} \frac{(N_{H_0, i} - N_{{\rm observed,} i})^2}{N_{H_0, i}} \eqstop}

For each choice of $e_{\rm cut}$, we construct $\chi^2$ as a function of $\mathcal{R}$ as shown in \reffig{chi2}. We then calculate the $\chi^2$ corresponding to a P-value of 0.05 for this data, indicated by the dashed line, with degrees of freedom, $\nu$, set by the number of frequency bins, $k$, in our observation. In this test, because the population parameters are known, and because the total number of counts across all bins is not fixed \textit{a priori}, we do not reduce the degrees of freedom below $k$. Performing a one-sided upper tail test for the Pearson $\chi^2$ statistic, values of $\chi^2$ above the P=0.05 threshold denote regimes where LISA can significantly reject the null hypothesis. We see that in nearly all cases, the ability to distinguish different populations remains driven by the lowest eccentricity cutoff of $e_{\rm cut}=0.01$. Referring back to ~\reffig{edist_counts}, we see that this conclusion seems driven by the $e_{\rm cut}=0.01$ counts appearing to change the most as we consider different populations. We note two conclusions from ~\reffig{chi2}. First, the first column of plots suggest that the two more eccentric distributions (In-Cluster, Galaxy Center) are significantly discernible even when present at substantially smaller abundances than the low eccentricity distribution. That is, LISA should be able to reasonably discern the presence of a highly eccentric sub-population without $e_{\rm cut}>0.01$ templates. Second, if we consider a population composed of a high eccentricity distribution (e.g. In-Cluster) there are still large regimes in the In-Cluster/Galaxy Center plots of ~\reffig{chi2} where LISA can discern between these highly eccentric populations driven by the lack of low eccentricity events expected from the Galaxy Center distribution. As a cautionary point, the accuracy of the Pearson chi-square goodness of fit test and our use of $P=0.05$ as a testing criteria degrades as the total number of observations in the dataset becomes small. When considering highly eccentric distributions as the model, with small choices of $e_{\rm cut}$, this condition is relevant and the significance of the corresponding statistical statements should be metered accordingly.  

\begin{figure*}[ht]
  \centering
  \includegraphics[width=\textwidth]{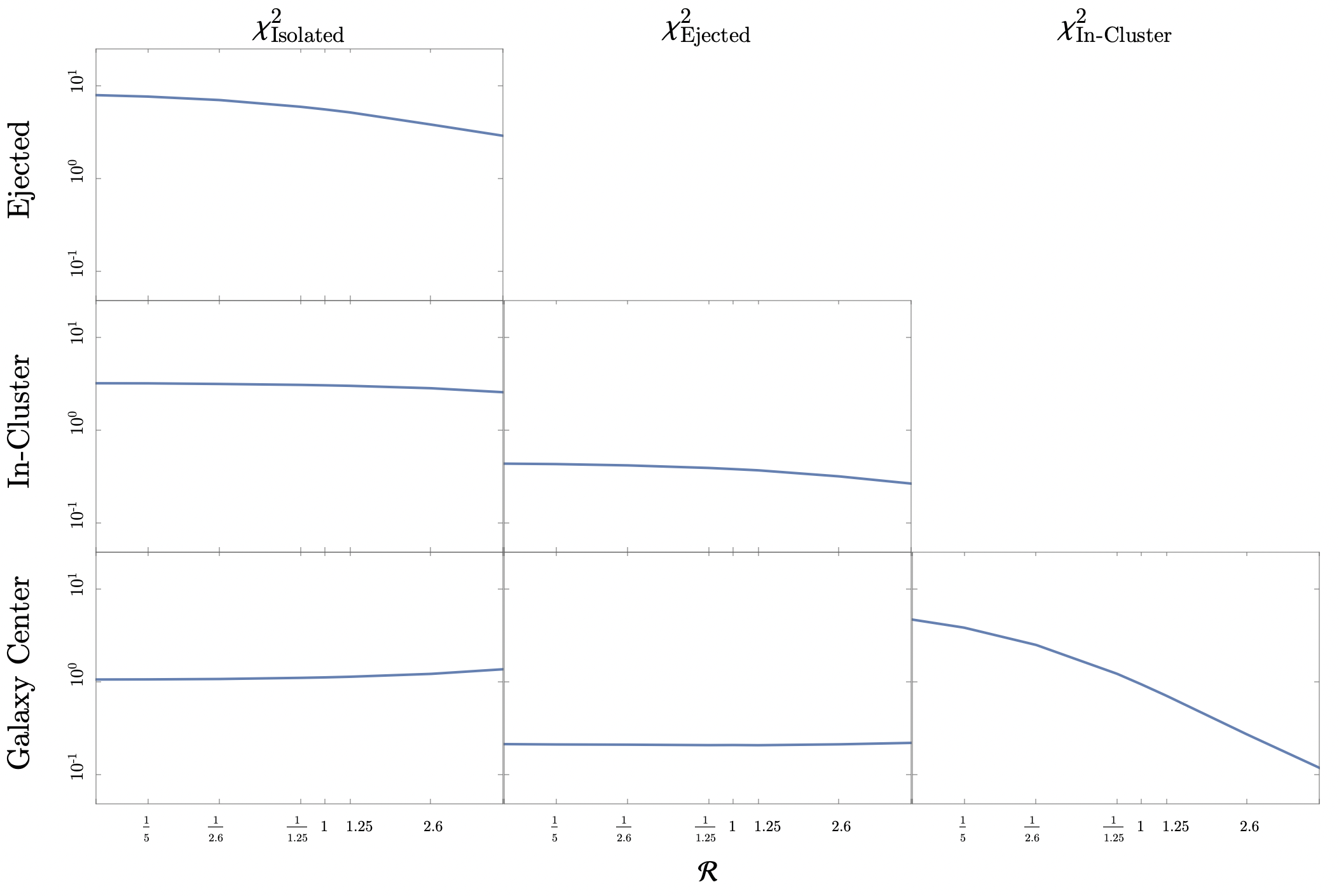}
  \caption{
  The ratio of the $\chi^2$ values corresponding to the difference in counts between $e_{\rm cut}=0.01$ and $e_{\rm cut}=0.1$ (blue line in \reffig{chi2_second}) to the $\chi^2$ values corresponding to the absolute counts for $e_{\rm cut}=0.01$ (red line in \reffig{chi2}). 
  }
  \label{fig:chi2_ratio}
\end{figure*}

To better explore the usefulness of $e_{\rm cut}>0.01$, we consider another observable - the difference in counts per frequency bin in LISA between an $e_{\rm cut}=0.01$ observation and an $e_{\rm cut}=0.1, 0.4$ or $0.9$ observation. Using this observable, we construct $\chi^2$ as above and the result of this study is shown in \reffig{chi2_second}. Two conclusions are worth noting. First, that having $e_{\rm cut}=0.1$ provides comparable  distinguishing power (i.e. an $\mathcal{O}$(1) effect on $\chi^2$) to the absolute counting with $e_{\rm cut}=0.01$ of \reffig{chi2}. There are specific cases when this observable greatly improves (i.e. an $\mathcal{O}(10)$ factor) upon $\chi^2$ over absolute counting with $e_{\rm cut}=0.01$: when distinguishing the lower-eccentricity distributions (e.g. Isolated versus Ejected, and Isolated versus In-Cluster); and, when distinguishing the most eccentric distributions (e.g. In-Cluster versus Galaxy Center) for $\mathcal{R}\lesssim 1$. Most prominently, this observable can discern the presence of an Ejected population, assuming an Isolated population as the model, over the full range of $\mathcal{R}$, which the absolute counting observable fails to do. In this case, access to $e_{\rm cut}>0.01$ makes a significant contribution to the physics reach of LISA. This result suggests that access to at least slightly larger ($e_{\rm cut}=0.1$) eccentricity templates may prove useful for distinguishing distributions which are both relatively low in eccentricity. Second, while having access to $e_{\rm cut}=0.4, 0.9$ generally outperforms $e_{\rm cut}=0.1$ by an $\mathcal{O}(1)$ factor, the improvement offered by the even higher $e_{\rm cut}=0.9$ is marginally different than that offered by $e_{\rm cut}=0.4$. This suggests that while there are modest improvements offered in going to $e_{\rm cut}>0.1$, there is little to gain in going to even higher $e_{\rm cut}>0.4$ values. To illustrate where extending to $e_{\rm cut}=0.1$ sensitivity offers valuable improvement over the $e_{\rm cut}=0.01$ case, we compute the ratio of $\chi^2$ for the difference in counts observable to that of the absolute counts observable in the lowest eccentricity cases in \reffig{chi2_ratio}.

One final point of interest concerns the ability of Hz range detectors like LIGO to follow up LISA observations of BBH events by detecting the merger of those which coalesce in a reasonable amount of time. In ~\reffig{edist_counts} we identified the subset of BBH pairs which will merge within a 10 year period. Note that all these events fall within the $e_{\rm cut} < 0.1$ threshold, and so should be expected to possess very little residual eccentricity at the time of their merger. We can add one additional contribution to our $\chi^2$ metric if we consider how many mergers would be observed in a follow-up detection by a Hz range detector. This amounts to comparing the $\tau < 10$ year counts between distributions. These counts approximately sum to 55 for the Isolated, 55 for the Ejected, 40 for the In-Cluster, and 27 for the Galactic Center distributions. This means that even for the most extreme values of $\mathcal{R}$, there would at most be a difference of about 28 merger events between distributions, corresponding to an additional contribution to $\chi^2$ of approximately $14$. Yet, we see in \reffig{chi2} that for $e_{\rm cut}=0.01$ that even in the worst case scenarios we have $\chi^2 \gtrsim 14$ and so we would still expect the distinguishing power to be generally driven by the mHz range observation.

\section{Conclusions} 
\label{section6}

We have shown that well-motivated populations of stellar mass binary black holes, characterized by their eccentricity distributions, can produce substantially different observations in gravitational wave detectors. For dHz range detectors, we have found: (1) due to the dynamics of binary evolution, the duration of gravitational wave emission at higher frequencies is relatively brief and we have shown that this limits the effectiveness of dHz range detectors at discerning binary populations of stellar mass black holes using eccentricity unless the sensitivity of these detectors is substantially increased with respect to that of modern mHz range detectors; and, (2) we have shown that a cluster of observations at high frequencies is effectively hidden from even high sensitivity dHz range gravitational wave detectors without access to very large ($e>0.4$) eccentricity gravitational waveform templates. In the case of mHz range detectors, we found: (3) the ability of gravitational wave detectors to observe at different eccentricity cutoffs was shown to lend information that breaks degeneracies between datasets corresponding to different relic populations of stellar mass binary black hole pairs;  (4) for reasonable relative abundances of eccentric populations, a detector like LISA can discern different stellar mass BBH populations with statistical significance using only $e_{\rm cut}=0.01$ templates; (5) while higher $e_{\rm cut}$ result in more observations, which improve errors, lower $e_{cut}$ can reveal discrepancies in the expected observation that are more statistically significant, especially between higher eccentricity distributions; (6) the difference in event counts for different choices of $e_{\rm cut}$ is a useful observable in addition to the absolute counts, outperforming the absolute counting in some cases; (7) the absolute counting observable is driven by $e_{\rm cut}=0.01$ while the difference-in-counts observable performs well for $e_{\rm cut}=0.1$, somewhat better for $e_{\rm cut}=0.4$, but its effectiveness is saturated for $e_{\rm cut}>0.4$ suggesting the incorporation of $e_{\rm cut}>0.4$ templates is unlikely to be useful for discerning eccentric stellar mass BBH populations; and, (8) the multi-channel observation of stellar mass BBH merger events and lack of corresponding observation at lower frequencies is a relevant signal, though statistically subdominant to the mHz frequency observation alone and, consequently, this work has primarily promoted the effectiveness of studying BBHs at the population level. This work argues that eccentricity is a useful probe of the formation channels of stellar mass black holes we observe in our Universe; it should be used as a tool in building our understanding of black hole processes and serve as a source of evidence for new theories of physics with connections to black holes.

\section{Acknowledgements}
\label{acknowledge}

ND is supported in this work by the Society of Fellows at Boston University and the Graduate Fellowship for STEM Diversity (National Physical Science Consortium Graduate Research Fellowship). ZX is supported by the National Key R\&D Program of China (2021YFC2203100), NSFC under Grant No.\ 12275146, an Open Research Fund of the Key Laboratory of Particle Astrophysics and Cosmology, Ministry of Education of China, and a Tsinghua University Dushi Program. The work of LR is supported by NSF Grant Nos. PHY-1620806 and PHY-1915071, the Chau Foundation HS Chau postdoc award, the Kavli Foundation grant “Kavli Dream Team,” and the Moore Foundation Award 8342.

\bibliography{bibliography.bib}

\appendix

\end{document}